# Electrical addressing of compensated magnetic order in artificial spin ice

Stéphane Nilsson[1,2*], Lauren J. Riddiford[1,2*], Emily Darwin[3,4], Hans J. Hug[3,4], Aleš Hrabec[1,2], Laura J. Heyderman[1,2]

[1]Department of Materials, Laboratory for Mesoscopic Systems, ETH Zurich, 8093 Zurich, Switzerland

[2]PSI Center for Neutron and Muon Sciences, 5232 Villigen PSI, Switzerland

[3]Swiss Federal Laboratories for Materials Science and Technology (Empa), 8600 Dübendorf, Switzerland

[4]Department of Physics, University of Basel, 4001 Basel, Switzerland

E-mail: stephane.nilsson@psi.ch, lauren.riddiford@psi.ch

## Abstract

Compensated magnetic systems, such as antiferromagnets, hold immense potential for spintronic devices thanks to vanishing stray fields, robustness to perturbations, and ultrafast dynamics. Yet, electrical writing and reading of their order remains a major challenge. Here we use artificial spin ice as a mesoscale platform in which compensated order can be engineered, directly imaged, and electrically controlled. Specifically, we demonstrate combined electrical writing and reading of a compensated state in a triangular artificial spin ice. Exploiting the orientation dependence of the spin–orbit torque switching current, we independently switch three nanomagnet sublattices through a single current channel, with the planar Hall effect providing sublattice-resolved readout. This grants access to all eight magnetization states of the triangular plaquette, including its fully compensated state, yielding a nonlinear, history-dependent electrical response. Our results establish an approach to electrically programmable compensated magnetic systems and provide an important foundation for reconfigurable magnetic devices and unconventional computing.

# Introduction

Compensated magnetic systems, such as antiferromagnets, offer an attractive platform for next-generation spintronic technologies. Unlike ferromagnets, which generate dipolar stray fields that can mediate cross-talk between neighboring devices and limit integration density, compensated systems have a vanishing net magnetization. This suppresses stray-field interactions and provides robustness against external magnetic field perturbations, while giving intrinsically fast spin dynamics that offer prospects for ultrafast information processing in the terahertz regime.

Despite these advantages, exploiting compensated magnetic order in functional devices remains challenging owing to the difficulty of electrical writing and readout of compensated magnetic states[1,2]. In general, deterministic control requires a mechanism that couples selectively to the staggered order parameter rather than to the net magnetization[3]. For example, electrical writing using spin–orbit torques (SOT) has been demonstrated in antiferromagnetic CuMnAs[3] and $Mn_2Au$[4], where the crystal symmetries enable staggered spin–orbit torques, and also in antiferromagnetic NiO[5] using a conventional damping-like torque. However, systematic investigation of the switching mechanisms is hindered by the fact that the switching is limited to nanoscopic domains that do not retain their state due to the strong magnetoelastic coupling[6,7]. Similarly, the absence of a net magnetic moment makes compensated states difficult to distinguish using conventional magnetometry[8]. Although specific non-collinear geometries[9] or altermagnetic systems[10] can yield measurable Hall signals, electrical readout of antiferromagnetic order remains a major open challenge.

Artificial spin ices (ASIs) provide a versatile mesoscale platform for engineering and manipulating magnetic order through lithographically defined arrays of interacting nanomagnets[11–13]. Previous approaches to controlling ASI configurations have exploited geometry engineering[14,15], global or local magnetic fields[16–20], and thermal relaxation achieved through global heating[21–23] or localized laser illumination[24–26]. A complementary range of techniques has been developed to probe these states, including directly imaging the magnetic states using magnetic force microscopy[27] and x-ray photoemission electron microscopy[15], and electrical approaches based on ferromagnetic resonance[28–30], magnetoresistance[31], and magnetic tunnel junctions[32]. However, simultaneous electrical writing and readout of the complete magnetic state space of a compensated ASI has not yet been realized.

In this work, we demonstrate the combined electrical writing and readout of all of the magnetic states in a triangular ASI, which comprises three coupled sublattices of single-domain nanomagnets (Fig. 1). The magnetization in each nanomagnet can be considered to be a macrospin pointing in one of two directions along the nanomagnet long axis, and the collective orientations of the macrospins give rise to a discrete set of magnetic configurations, including states with compensated magnetic order. By injecting current pulses of selected amplitude through an underlying heavy-metal layer, we achieve field-free, selective spin–orbit torque switching of the individual nanomagnet sublattices. This provides full access to all eight magnetization states of the triangular plaquettes, enabling the controlled writing and electrical readout of fully compensated magnetic configurations with zero net magnetization. The magnetic configurations are independently confirmed with spatial imaging, allowing the electrical readout to be directly correlated with the underlying magnetic order.

Furthermore, by tuning the angle of the probing magnetic field during planar Hall readout, contributions from individual or multiple sublattices can be selectively detected. This tunable readout produces a rich, history-dependent, and nonlinear electrical response and, combined with tailored current pulses that reconfigure the magnetic state, provides two key functions required for reconfigurable ASI and physical reservoir computing architectures[31,33–35].

# Results

## Electrical writing and detection of magnetization switching

The device layout used in this work is schematically illustrated in Fig. 1a, with scanning electron microscopy (SEM) images of a fabricated device and of the triangular ASI shown in Figs. 1b and 1c, respectively. The arrays of nanomagnets, made of CoFeB (2.3 nm) / MgO (1.2 nm) / Ta (2 nm), with lateral dimensions of 350 nm × 100 nm and a lattice parameter $a$ = 700 nm corresponding to the side length of the triangles, were patterned on top of Ta (6 nm) / W (2.2 nm) Hall bars. The angles and inputs used throughout this work are illustrated in Fig. 1a and defined as follows: $\theta_{nm}$ denotes the orientation of the nanomagnet long axis, $\theta_H$ the orientation of the probing magnetic field $\boldsymbol{H}_{probe}$, and $\boldsymbol{J}$ the current density applied for SOT-induced magnetization switching and readout, with all angles measured with respect to the $x$-axis. The nanomagnet arrays have dimensions of approximately 4 μm × 4 μm and are positioned at the center of the channel (Fig. 1b) in a region of uniform current, which is verified with finite-element simulations (Supplementary Note 1).

The SOT-induced magnetization switching events are electrically detected using differential planar Hall effect measurements[36,37]. The planar Hall resistance $R_H$ as a function of an applied magnetic field, measured at $\theta_H = 20^{\circ}$, is shown in Fig. 1d for nanomagnets at $\theta_{nm} = 90°$ with their magnetization initially oriented along $-y$ ($+y$) for the blue (red) curve. The signal reaches a minimum at $\mu_0 H = +6$ mT ($-6$ mT), corresponding to the threshold magnetic field beyond which magnetization reversal occurs. We therefore set the probing field amplitude to $|\mu_0 H_{probe}| = 5$ mT for every measurement, which is the largest field that can be applied before magnetic field-induced magnetization switching takes place. In the regime between $\pm H_{\text{probe}}$, the Hall resistance is linear with the applied magnetic field and, crucially, the slopes are of opposite sign for the two different initial magnetization directions of the nanomagnet. To exploit the difference in sign between the two states, we define the differential planar Hall resistance as:

$$\Delta R = R_{+H_{\text{probe}}} - R_{-H_{\text{probe}}} \tag{1}$$

which gives $\Delta R_{\downarrow} < 0$ and $\Delta R_{\uparrow} > 0$ for the two opposite initial magnetization states of the nanomagnet, as indicated in Fig. 1d. Thus, by recording $\Delta R$ as a function of current pulse amplitude, magnetization switching events can be identified.

The evolution of $R_H$ as a function of applied magnetic field, measured at various magnetic field angles $\theta_H$, is presented in Fig. 1e for nanomagnets oriented at $\theta_{\text{nm}} = -20°$ (see Supplementary Figs. 2a and 2b for $\theta_{\text{nm}} = -80°$ and $\theta_{\text{nm}} = 40°$, respectively). The $|\Delta R|$ is largest at $\theta_H \approx -90°$ and decreases as $\theta_H$ is increased, reaching zero when the magnetic field is aligned parallel to the long axis at $\theta_H = -20°$. Furthermore, $\Delta R$ undergoes a sign reversal upon passing through $\theta_H = -20°$, with $R_H$ in Fig. 1e exhibiting positive peaks for $\theta_H = -30°$ and negative peaks for $\theta_H = -10°$ on sweeping the magnetic field. As a direct consequence, the $\Delta R$ versus current loops shown in Fig. 1f display opposite polarities for $\theta_H = -30°$ and $\theta_H = -10°$. This angular dependence of the planar Hall response is well reproduced by micromagnetic simulations, as shown in Supplementary Note 2.

Now that we have shown how to measure the direction of the magnetization in the nanomagnets at different magnetic field angles with respect to the current, we implement the following two-step sequence: (i) a 10 ns write current pulse is first applied through the current channel to drive SOT-induced magnetization switching of the nanomagnets, then (ii) to detect SOT-induced switching, $\Delta R$ is extracted by measuring $R_H$ under a 200 µA read current applied for 200 ms at $\pm H_{probe} = \pm 5$ mT. By repeating this sequence for increasing write current pulse amplitudes, hysteresis loops of $\Delta R$ as a function of applied current amplitudes are obtained, as shown in Fig. 1f. These hysteresis loops provide clear evidence of SOT-induced magnetization reversal in single-orientation nanomagnet arrays, examples of which are shown in the SEM images of Figs. 2a and 2b. In agreement with the results shown in Fig. 1e, the $\Delta R$ signal amplitude shown in Fig. 1f varies with the probing magnetic field orientation and the loop polarity reverses sign upon crossing $\theta_H = -20°$.

The mean critical switching current and the switching current distribution have been extracted from SOT-induced switching loops for arrays of nanomagnets with a single orientation for several different $\theta_{nm}$. Example $\Delta R$ versus current loops are shown for nanomagnet orientations of $\theta_{nm} = -80°$, $40°$, and $-20°$ in Fig. 2c, with the shaded areas indicating the switching current distribution among individual nanomagnets. The evolution of the critical switching current as a function of nanomagnet orientation $\theta_{nm}$ is presented in Fig. 2d, with the error bars representing the switching current distribution and the shaded areas highlighting the distribution for the three nanomagnet orientations in the triangular lattice, namely $\theta_{nm} = -80°$, $40°$, and $-20°$. Two regimes can be identified: (i) in the range $0° < \theta_{nm} < 50°$, the critical switching current exhibits a strong dependence on $\theta_{nm}$, more than doubling from 22 mA to 50 mA when decreasing the angle from $\theta_{nm} = 50°$ down to $\theta_{nm} = 10°$; (ii) in the range $50° < \theta_{nm} < 90°$, the critical switching current shows only a weak dependence on orientation, remaining relatively constant across this regime. These two regimes reflect a fundamental change in the switching mechanism[38]: for nanomagnets almost perpendicular with respect to the current channel ($\theta_{nm} \rightarrow 90°$, Type Y switching), magnetization reversal is driven by precessional motion, which is an efficient switching pathway requiring relatively low current densities. When the nanomagnet long axis approaches alignment with the current ($\theta_{\mathrm{nm}} \rightarrow 0°$, Type X switching), the magnetization reversal occurs without precession and the spin-orbit torques need to directly overcome the energy barrier associated with the nanomagnet shape anisotropy, resulting in the sharp increase of the critical switching current observed experimentally. The measured angular dependence is reproduced by macrospin simulations (see Supplementary Note 3) and is consistent with predominantly damping-like spin-orbit torques as presented in Supplementary Fig. 4. Furthermore, the Oersted field at the maximum switching current remains below the nanomagnet coercive field, indicating that it is not responsible for the magnetization switching (Supplementary Fig. 1b).

**Selective current-induced switching of sublattices**

The strategy to achieve selective switching of individual nanomagnet sublattices in the triangular lattice using a single current channel relies on the strong dependence of the critical switching current on $\theta_{\mathrm{nm}}$ established above. For this, the triangular lattice is oriented such that $\theta_{\mathrm{nm}} = -80°$, 40°, and −20° for the three sublattices, which are colored in blue, pink, and green respectively in Fig. 1a. This ensures that there is no overlap of critical switching currents between the sublattices, as highlighted by the shaded regions in Figs. 2c and 2d, enabling sequential and independent SOT-induced switching of each sublattice employing a single current channel.

To build toward the triangular lattice, we first consider arrays of nanomagnets with two distinct sublattices. SEM images of these nanomagnet arrays, comprising the three different combinations of nanomagnet pairs, are presented in Supplementary Note 4. For each nanomagnet array, $\Delta R$ versus current loops were recorded at various $\theta_H$, and the results for one combination of nanomagnet pairs are presented in Fig. 3. At $\theta_H = 20°$ (top plot), the $\Delta R$ versus current loop exhibits contributions from both sublattices, i.e. the nanomagnets at $\theta_{\mathrm{nm}} = -80°$ (shown in blue) and those at $\theta_{\mathrm{nm}} = -20°$ (shown in green). Four distinct jumps in $\Delta R$ are observed, each corresponding to the switching of one sublattice, as indicated by the numbered labels ① to ④ and illustrated schematically. Because of the well-defined separation in critical switching currents between the sublattices, the reversal of each sublattice can be unambiguously identified. Similarly, the switching events of each sublattice can also be resolved in the nanomagnet arrays with $\theta_{\mathrm{nm}} = -80°$ and 40°, and $\theta_{\mathrm{nm}} = 40°$ and −20° (see Supplementary Fig. 6).

Each of the two sublattices can also be individually resolved. At $\theta_H = -80°$, shown in the middle plot of Fig. 3, only two jumps in $\Delta R$ are observed, corresponding to the switching of the $\theta_{\mathrm{nm}} = -20°$ sublattice. Although the $\theta_{\mathrm{nm}} = -80°$ sublattice also undergoes magnetization reversal, its $\Delta R$ contribution vanishes at this $\theta_H$, as established in Fig. 2e, and the switching events in this sublattice are not detected. Conversely, at $\theta_H = -20°$ (bottom plot), the $\theta_{\mathrm{nm}} = -20°$ sublattice exhibits a vanishing $\Delta R$ contribution, and only the switching events of the $\theta_{\mathrm{nm}} = -80°$ sublattice are detected.

The results for nanomagnet arrays with two sublattices demonstrate three key points. First, the dependence of $\Delta R$ on $\theta_H$ observed for single-orientation nanomagnet arrays is preserved in two-orientation nanomagnet arrays, confirming that each sublattice contributes independently to the measured signal. Second, with appropriate selection of $\theta_H$, the switching events of either one or both sublattices can be selectively detected. Third, the non-overlapping critical switching current ranges of the two sublattices ensure that their individual switching events can always be unambiguously distinguished. As will be shown in the following section, these properties extend to the triangular lattice comprising all three nanomagnet sublattices.

**Electrical writing and readout of a compensated state**

Building on the results obtained for the two-sublattice arrays of nanomagnets, we now apply the same approach to the triangular ASI and demonstrate access to all $2^3 = 8$ magnetization states, labeled ① to ⑧ in Figs. 4b and 4c. In the top plot of Fig. 4a ($\theta_H = -60°$), the write-pulse amplitude is swept from −46 mA to 46 mA; this major loop runs through six of the eight states: ① → ② → ③ → ④ as the $\theta_{\mathrm{nm}} = -80°$, 40° and −20° sublattices are successively reversed by positive pulses, and ④ → ⑤ → ⑥ → ① as they are reversed back in the same order by negative pulses. States ⑦ and ⑧ are reached with minor loops in which the $\theta_{\mathrm{nm}} = -20°$ sublattice is not reversed: in the middle plot of Fig. 4a ($\theta_H = -60°$), state ⑦ can be achieved by a write-pulse sweep from −46 mA to 36 mA, running through states ① → ② → ③ with positive current pulses, and through states ③ → ⑦ → ① with negative current pulses. State ⑧ is similarly obtained with a minor-loop current sweep from 46 mA to −36 mA. The complete current pulse protocols required to access all eight magnetization states are illustrated in Figs. 4b and 4c. The major and minor loops thus differ only in their maximum pulse amplitude ($\theta_H = -60°$ in both cases), but yield markedly different $\Delta R$ responses, as $\Delta R$ at a given current pulse amplitude depends on the preceding pulse sequence, leading to the history dependence of the signal. At $\theta_H = 90°$ (bottom plot of Fig. 4a), all three switching events are again resolved, but with a clearly different $\Delta R$ response from $\theta_H = -60°$, further demonstrating that the electrical readout of the triangular ASI can be tuned by the orientation of the probing magnetic field, leading to a rich, non-linear electrical response. The full orientation-dependent $\Delta R$ versus current plots for the triangular lattice are presented in Supplementary Note 5.

Magnetic force microscopy (MFM) frequency shift images of the triangular ASI following different current pulse protocols are shown in Fig. 4d. The MFM image on the left corresponds to the magnetic state of the nanomagnet array after the application of a large positive current pulse so that all three sublattices have been switched simultaneously, driving the system into a ferromagnetic-like state with the net magnetization oriented along -y, corresponding to state ④. The magnetic configuration ⑤ is displayed in the righthand image, which was obtained by applying a large positive current pulse followed by a small negative current pulse as presented in Fig. 4c. This protocol drives the system into the magnetic ground state of the triangular ASI, which is characterized by alternating flux-closed loops of spins in the triangular plaquettes (Fig. 4e). Here a small number of nanomagnets oriented at $\theta_{nm}$ = 40° were unintentionally reversed as a result of a slightly excessive pulse amplitude. Notably, this transition from an uncompensated to a fully magnetically compensated state is achieved entirely through current pulse protocols, without the need for a staggered conjugate field. Despite the vanishing net moment, the planar Hall measurement still yields a clear, non-vanishing $\Delta R$ signal for this compensated magnetic configuration, whose amplitude can be further tuned by varying $\theta_H$ (Fig. 4a). This constitutes a direct electrical readout of a magnetically compensated state, a capability that remains one of the central open challenges in antiferromagnetic spintronics, where the absence of net magnetization typically renders conventional magnetoresistive readout ineffective.

## Discussion

In summary, we have demonstrated full electrical writing and readout of individual sublattices in a triangular ASI, utilizing spin–orbit torques for manipulation and the planar Hall effect for detection. By orienting the lattice geometry to give non-overlapping critical switching current windows, all eight magnetization states of the triangular plaquettes, including the fully compensated magnetic state, can be selectively addressed and read through a single current channel.

Importantly, the fully compensated state produces a finite planar Hall response despite its zero net magnetization, reminiscent of recent observations of Hall responses in compensated altermagnetic systems[10]. Our artificial spin ice therefore provides a versatile mesoscale platform for studying the electrical response of engineered compensated magnetic order and its dependence on the magnetic configuration.

While the flux-closure configuration of the compensated state is suggestive of toroidal ordering, the structural symmetry of our triangular lattice prevents it from possessing a net toroidal moment[39], which fundamentally requires the simultaneous breaking of spatial inversion and time-reversal symmetries. Instead, a route to access toroidal order lies in the approach implemented in Ref. [20]. Specifically, if we replace each nanomagnet in the triangular unit cell with a pair of side-by-side nanomagnets, this would allow antiferrotoroidal order to be established through the same current-pulse protocol, with the two nanomagnets in each pair having the same magnetization direction. While controlling toroidal order typically necessitates conjugate, combined magnetic and electric fields[40], with our approach SOT-driven sublattice switching would provide a streamlined, purely electrical route to address such states.

Furthermore, the combination of multiple accessible spin states, nonlinear switching dynamics, and history-dependent resistance response fulfills the key prerequisites for physical reservoir computing. This methodology readily extends to other ASI geometries, such as kagome[41] or pinwheel lattices[18], where tuning nanomagnet interactions, orientation, and magnetic frustration can further enrich the accessible state space for advanced neuromorphic computing and reconfigurable magnonics.

# Methods

### Device fabrication

The devices were fabricated by first depositing Ta (6 nm) / W (2.2 nm) / CoFeB (2.3 nm) / MgO (1.2 nm) / Ta (2 nm) onto a Si substrate coated with an insulating $Si_3N_4$ layer using DC magnetron sputtering for the metallic layers and RF magnetron sputtering for the MgO layer. The deposition was carried out at a base pressure of $<5.5 \times 10^{-8}$ mbar and a deposition Ar pressure of $4 \times 10^{-3}$ mbar. Stadium-shaped nanomagnets with lateral dimensions of 350 nm × 100 nm were subsequently patterned by electron-beam lithography combined with Ar ion etching, with the etch stopping at the top of the W layer. For this, we employed a 15 nm-thick Al hard mask that defines the nanomagnets during the Ar ion etching. The hard mask was created by first exposing the pattern in an 80 nm-thick layer of poly(methyl methacrylate) (PMMA) followed by development in methyl isobutyl ketone and isopropanol (MIBK:IPA) 1:3 developer for 45 s. A 15 nm-thick Al layer was subsequently deposited with thermal evaporation at a base pressure of $5 \times 10^{-6}$ mbar. This was followed by lift-off in acetone. The Hall cross structures, comprising a current channel of 6 μm width and Hall probe arms of 1.5 μm width, were patterned in the Ta (6 nm) / W (2.2 nm) conducting bilayer with photolithography and Ar ion etching. For this, a 1 μm-thick S1813 positive tone photoresist was used as the hard mask. Finally, Cr (3 nm) / Au (100 nm) electrodes were defined by photolithography and deposited by thermal evaporation. The distance between the electrodes is 12 μm and 14 μm for the current channel and voltage arm electrodes, respectively.

### Measurements

All transport measurements were performed with the sample mounted on a rotating platform placed between the poles of an electromagnet, which could supply magnetic fields up to 800 mT. We delivered 10 ns square write current pulses using an AVTECH pulse generator and measured the Hall resistance using a Keithley 2182A nanovoltmeter with an integration time of 10 power line cycles (=200 ms) to obtain a time-averaged reading. A Keithley 6221 DC current source supplied the 200 μA read current, and a bias-tee prevented leakage of the current from the DC source into the pulse generator.

To measure the Hall resistance as a function of applied magnetic field, $R_H$, we first set the magnetic field to the desired value and waited 250 ms for the field to stabilize. We then applied a 200 μA read current $I$ and recorded the voltage $V$ across the Hall bar arms, from which we computed $R_H = V/I$.

To measure $\Delta R$, we first applied a 10 ns write current pulse, after which we measured the Hall resistance at +5 mT and −5 mT, denoted $R_{+5mT}$ and $R_{-5mT}$, respectively. We then computed the differential Hall resistance as $\Delta R = R_{+5mT} - R_{-5mT}$.

**Magnetic force microscopy**

Magnetic force microscopy (MFM) measurements were performed in a custom high-vacuum setup operating at a pressure of approximately $1 \times 10^{-6}$ mbar. Working under vacuum increases the mechanical quality factor of the cantilever to values as high as $Q \approx 1.5 \times 10^6$, providing substantially higher sensitivity than MFM measurements carried out in air. This increased sensitivity also makes it possible to use a very thin magnetic coating on the MFM probe, thereby reducing the influence of the tip on the magnetic state of the nanomagnets[42].

An initially uncoated SS-ISC silicon cantilever from Team Nanotech GmbH was used. The cantilever was 450 µm long and 35 µm wide, with a nominal tip radius below 5 nm. Its resonance frequency was experimentally determined to be 18.26 kHz. From the measured resonance frequency and the cantilever dimensions, a spring constant of 0.2 N/m was obtained[43]. To provide magnetic sensitivity, the cantilever was then coated by sputter deposition. A 2 nm-thick Ta layer was first deposited as a seed layer, followed by a 7 nm-thick Co magnetic layer.

During the measurements, the cantilever was driven at resonance using a phase-locked loop system from Zurich Instruments, while its oscillation amplitude was maintained at 10 nm. Magnetic interactions between the tip and the sample alter the force gradient experienced by the cantilever, resulting in shifts of its resonance frequency. These frequency shifts were recorded as the MFM signal. The measurements were conducted in non-contact mode, with capacitive feedback used to regulate the tip-sample separation[44]. The scan time per line was set to be 2 s.

**Acknowledgements.** The authors acknowledge financial support from the Swiss National Science Foundation (Grant No. 200020_200332). L.J.R. acknowledges funding from the ETH Zurich Postdoctoral Fellowship Program 22-2 FEL-006. E.D. and H.J.H. were supported through the SNF Lead Agency project 200021E_211828. The authors thank the cleanroom staff at the Laboratory for Nano and Quantum Technologies at the Paul Scherrer Institute for technical support. The authors also thank Manfred Fiebig and Nora Taufertshöfer for helpful discussions.

**Conflicts of Interest.** The authors declare no conflicts of interest.

**Data availability.** The data that support the findings of this article are openly available at [45].

**Authors contributions.** S.N., L.J.R., A.H. and L.J.H. conceived the work. S.N. fabricated the samples, performed the transport measurements with support from L.J.R., and carried out the simulations; S.N. analysed the data with help from L.J.R. and A.H; E.D. performed the MFM measurements with support from H.J.H; S.N., L.J.R., A.H. and L.J.H. worked on the manuscript together. All authors discussed the results and commented on the manuscript.

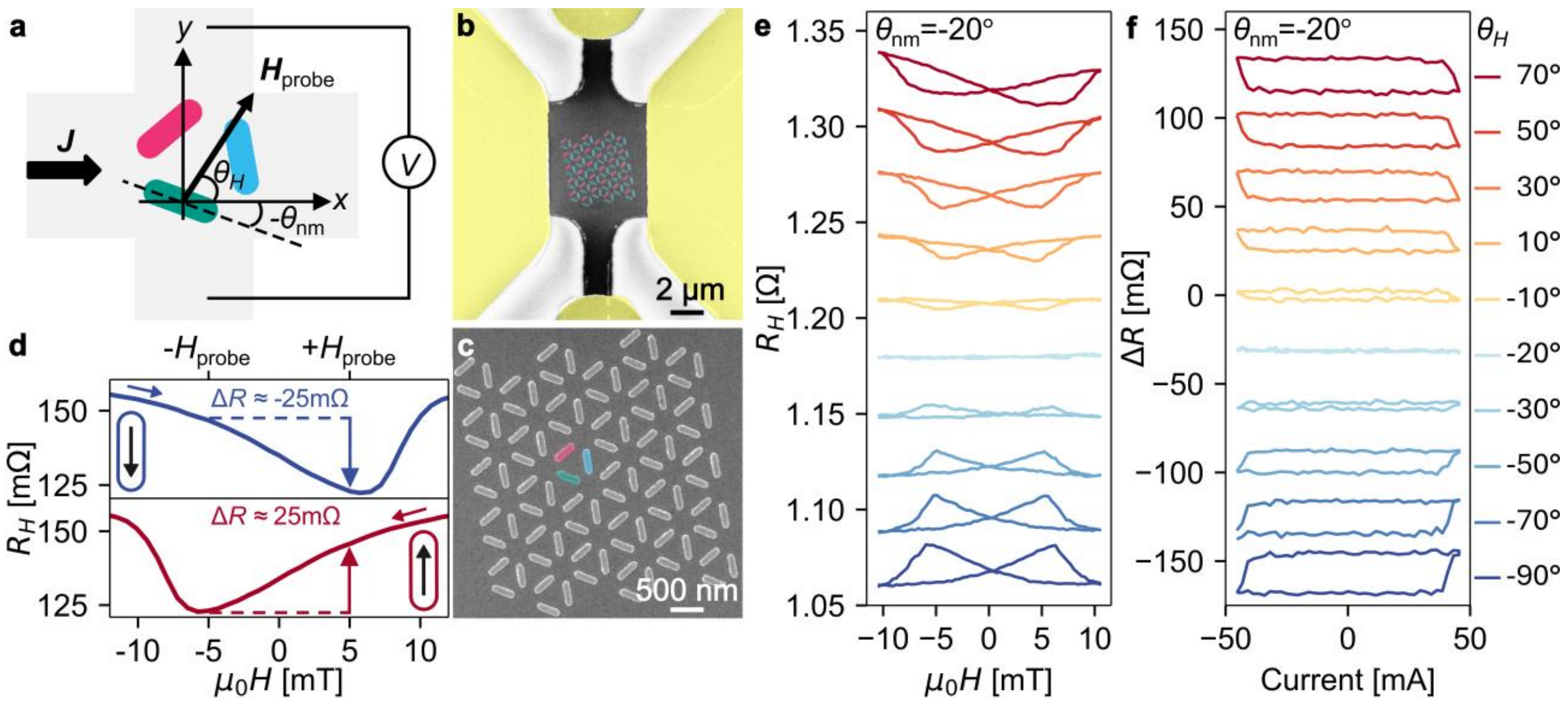

**Fig. 1: Electrical detection and control of the magnetic states in an array of nanomagnets**. (a) Schematic of the device, illustrating the orientation of the nanomagnet long axis $\theta_{\mathrm{nm}}$, and the orientation of the probing magnetic field $\theta_H$ with respect to the direction of the current $\boldsymbol{J}$, which is along $x$. Nanomagnets oriented at $\theta_{\mathrm{nm}}$ = -80°, 40°, and -20° are shown in blue, pink, and green, respectively. (b) Colorized scanning electron microscope (SEM) image of one of the fabricated devices with a triangular ASI, with the gold electrodes colored in yellow, the nanomagnets colored according to their respective orientation and the Hall cross with a dark grey contrast. (c) SEM image of the nanomagnets arranged in a triangular lattice, with each nanomagnet orientation for a single triangular plaquette highlighted in a different color. (d) Planar Hall resistance $R_H$ as a function of applied magnetic field $\mu_0 H$ for the nanomagnet magnetization initially oriented along $-y$ ($+y$) for the blue (red) curve in the top (bottom) subplot. The probing field amplitude, highlighted by the dashed lines, is set to $\pm\mu_0 H_{\mathrm{probe}} = \pm 5$ mT for each measurement. The corresponding $\Delta R$ signal amplitude is indicated by the dashed arrow. (e) $R_H$ as a function of applied magnetic field and (f) $\Delta R$ as a function of applied current for various probing field angles $\theta_H$, for nanomagnets oriented at $\theta_{\mathrm{nm}} = -20°$.

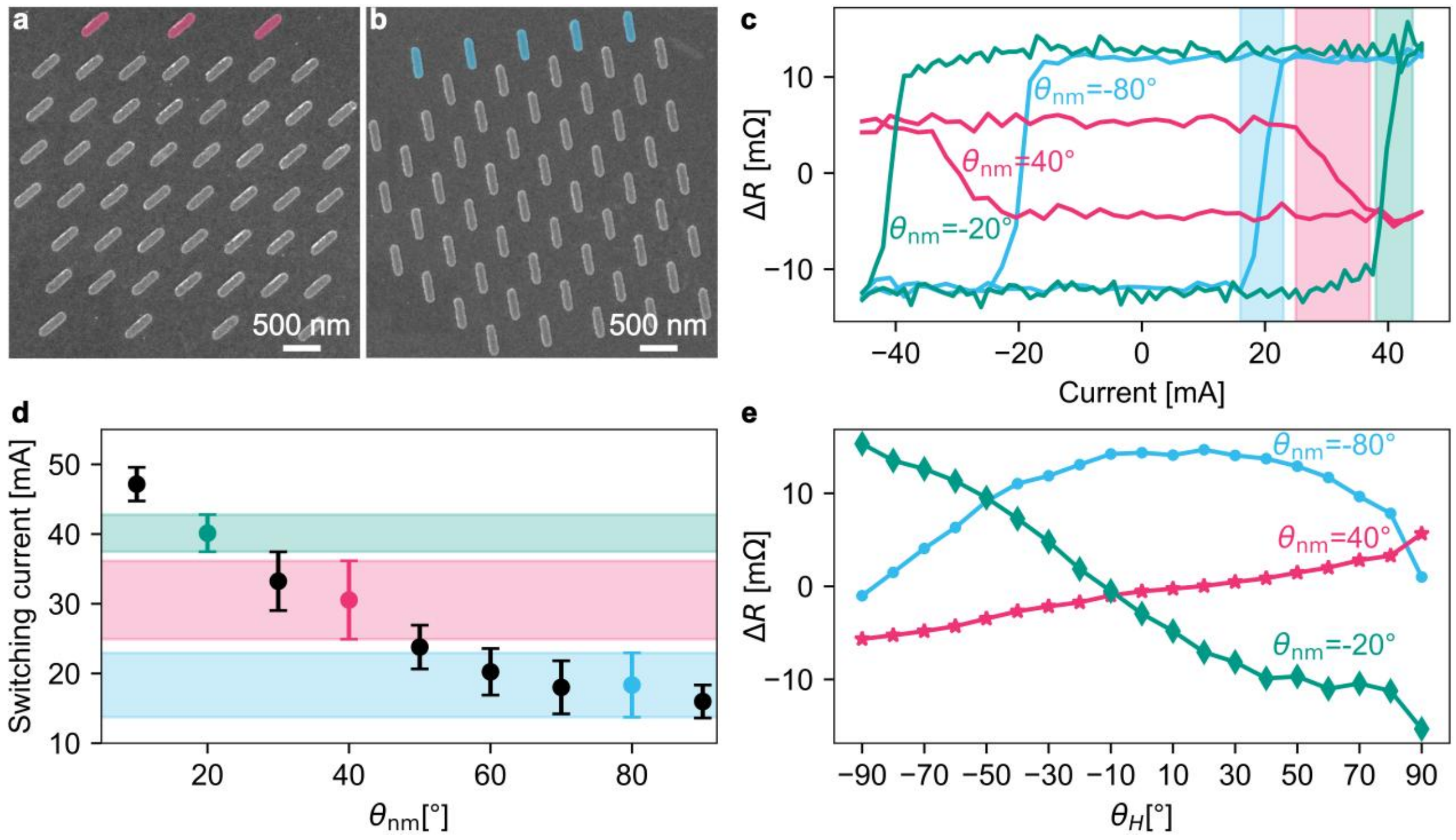

**Fig. 2: Dependence of the critical switching current and Δ*R* signal amplitude on nanomagnet orientation**. (a), (b) Colorized SEM images of arrays of single-orientation nanomagnets with their long axis oriented at (a) $\theta_{nm} = 40^{\circ}$ and (b) $\theta_{nm} = -80^{\circ}$. (c) Δ*R* versus current loops for nanomagnet arrays oriented at $\theta_{nm}$ = -80°, 40°, and −20°, with $\boldsymbol{H}_{probe}$ applied at $\theta_H$ = 10°, -90°, and -90°, respectively. The shaded regions indicate the extent of the switching current distribution for each orientation. (d) Critical switching current as a function of nanomagnet orientation $\theta_{nm}$, with the error bars indicating the range over which magnetization switching occurs and the shaded regions highlighting the range of the magnetization switching for the nanomagnets oriented at $\theta_{nm}$ = -80°, 40°, and −20°. (e) Δ*R* as a function of the probing magnetic field angle $\theta_H$ for arrays of nanomagnets with three different orientations. The colors in all panels are for data corresponding to the nanomagnets oriented at $\theta_{nm}$ = -80° (blue), 40° (pink), and −20° (green).

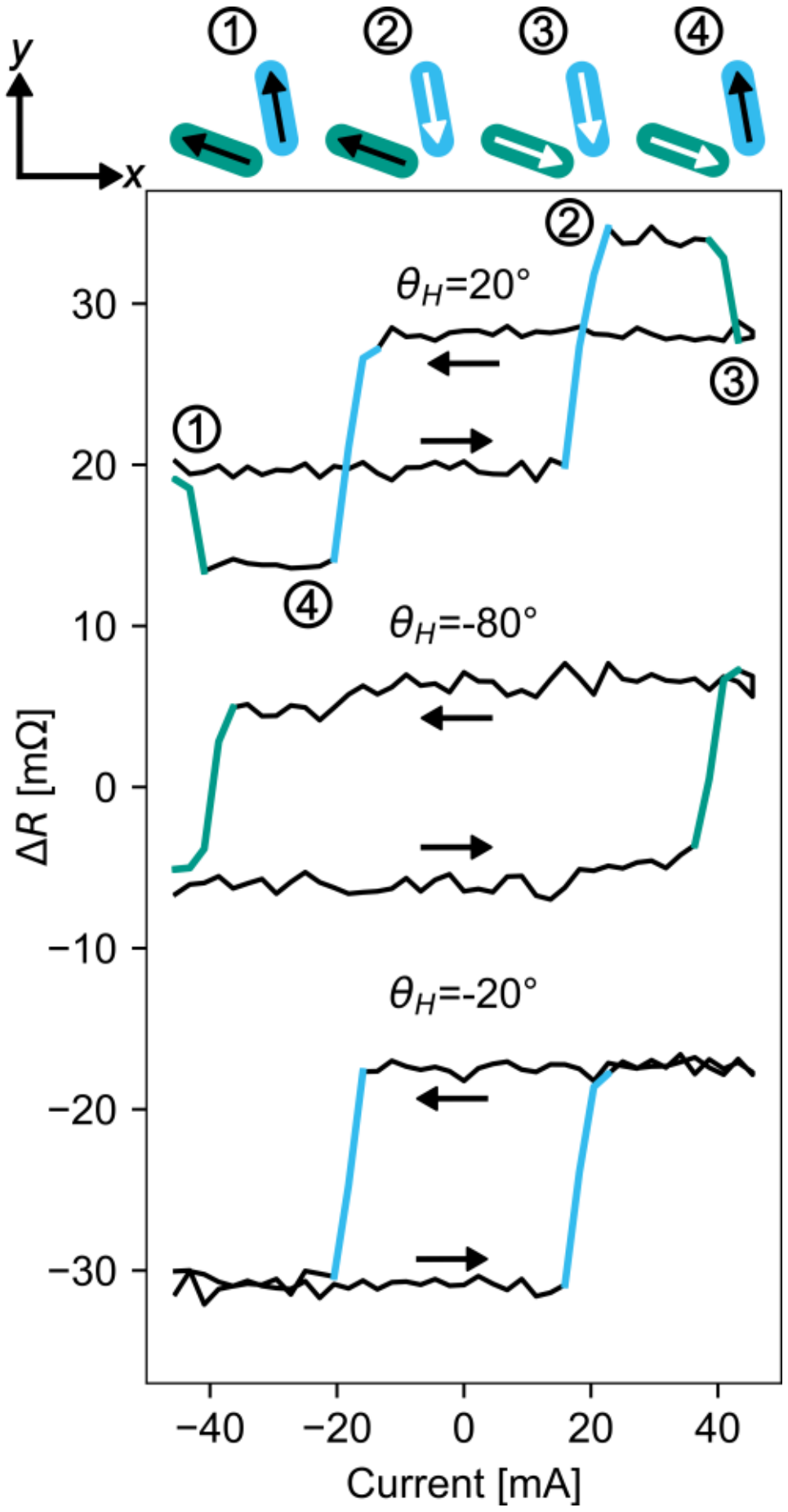

**Fig. 3: Tunable electrical readout of SOT-induced switching in an array of nanomagnets with two nanomagnet sublattices**. $\Delta R$ as a function of applied current pulse amplitude, measured at various $\theta_H$, with the nanomagnets oriented at $\theta_{nm} = -80^{\circ}$ and $-20°$. The switching events of the nanomagnets oriented at $\theta_{nm} = -80^{\circ}$ and $-20^{\circ}$ are highlighted in blue and green, respectively. The resulting magnetization state is schematically illustrated before and after each switching event, with the black and white arrows on the nanomagnets indicating the magnetization pointing towards $+y$ and $-y$, respectively. The black arrows next to the curves indicate the current sweep direction.

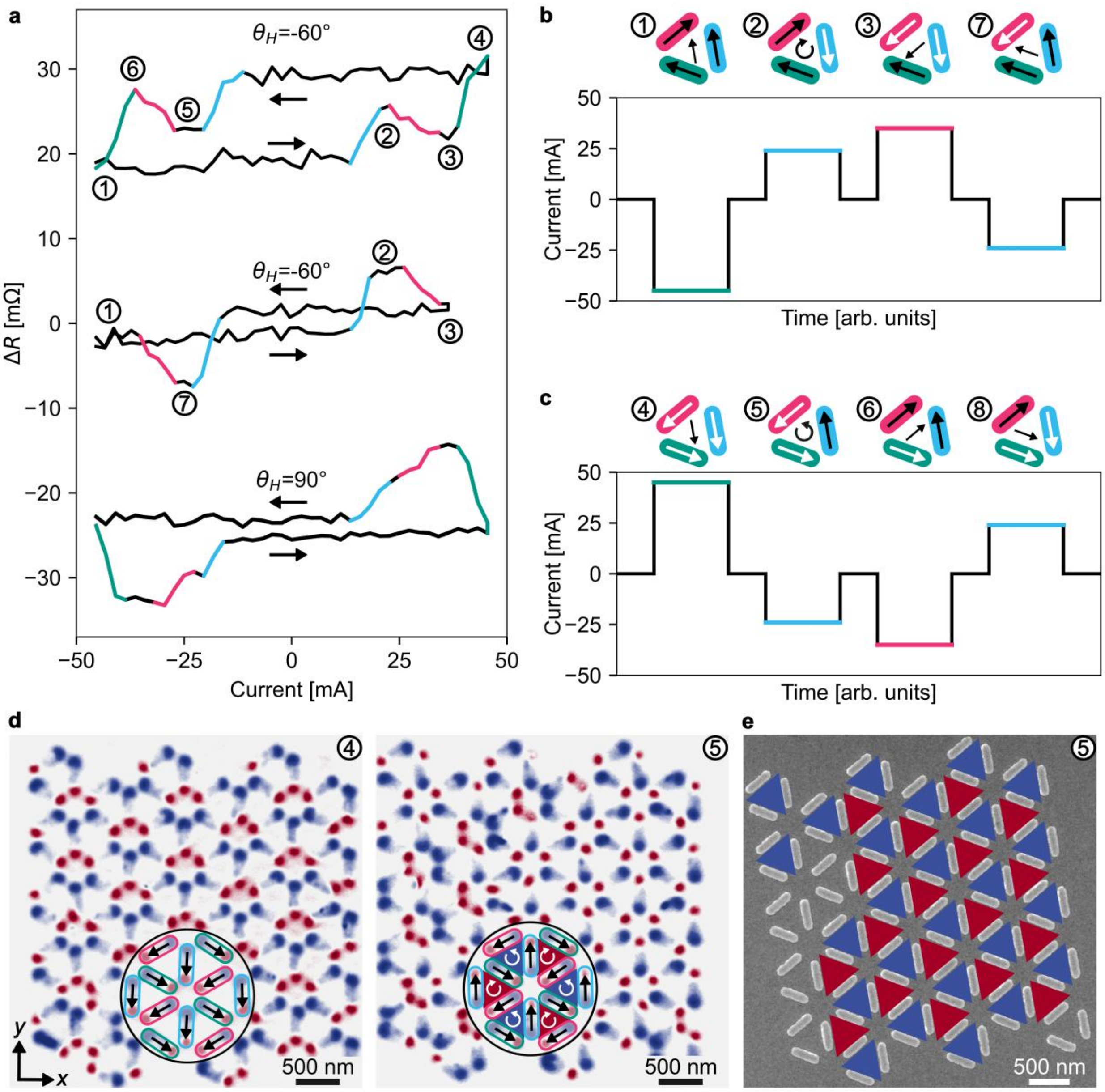

**Fig. 4: Electrical writing and reading of uncompensated and compensated states in the triangular ASI with three sublattices**. (a) $\Delta R$ as a function of applied current pulse amplitude, measured at various probing field angles $\theta_H$ for the triangular ASI. The switching events of the nanomagnets oriented at $\theta_{\text{nm}} = -80^\circ$, $40^\circ$, and $-20^\circ$ are highlighted in blue, pink, and green, respectively. (b)-(c) Current pulse protocols demonstrating sequential access to all eight magnetization states of the triangular plaquette. The magnetization state of the triangular plaquette is schematically illustrated after each switching event. The net magnetization is indicated by the arrow in the middle of each triangular plaquette, with the two opposite flux-closure states indicated with clockwise and anticlockwise arrows. (d) Magnetic force microscopy images of the array of nanomagnets following distinct current pulse protocols, depicting state ④ on the left and state ⑤ on the right; encircled regions highlight local nanomagnet orientations. (e) Scanning electron microscope image of the triangular ASI in state ⑤, with red and blue shaded triangular plaquettes indicating the opposing flux-closed magnetic configurations

# Supplementary information

# for

## Electrical addressing of compensated magnetic order in artificial spin ice

Stéphane Nilsson[1,2*], Lauren J. Riddiford[1,2*], Emily Darwin[3,4], Hans J. Hug[3,4], Aleš Hrabec[1,2], Laura J. Heyderman[1,2]

[1]Department of Materials, Laboratory for Mesoscopic Systems, ETH Zurich, 8093 Zurich,Switzerland

[2]PSI Center for Neutron and Muon Sciences, 5232 Villigen PSI, Switzerland

[3]Swiss Federal Laboratories for Materials Science and Technology (Empa), 8600 Dübendorf, Switzerland

[4]Department of Physics, University of Basel, 4001 Basel, Switzerland

E-mail: stephane.nilsson@psi.ch, lauren.riddiford@psi.ch

## Supplementary Note 1. Finite element method simulations

To determine the current density and Oersted field distributions within the Hall bar, finite-element method simulations were performed using the Magnetic and Electric Fields interface in COMSOL, with the results presented in Supplementary Fig. 1. The simulations were carried out for the steady-state case at the largest current required to fully switch the nanomagnets, namely 48 mA.

As shown in Supplementary Fig. 1a, current crowding occurs near the intersection of the current channel with the voltage probe arms. However, the nanomagnet array is positioned sufficiently far from this region, so that the current density across the nanomagnet array remains largely homogeneous.

The simulated Oersted field distribution is shown in Supplementary Fig. 1b and does not exceed 3.5 mT across the nanomagnet array. This value is well below the nanomagnet coercive field of approximately 6 mT, determined from the measurements presented in Fig. 1d, which means that field-induced magnetization reversal due to Oersted fields can be ruled out across the full range of current densities used in this work.

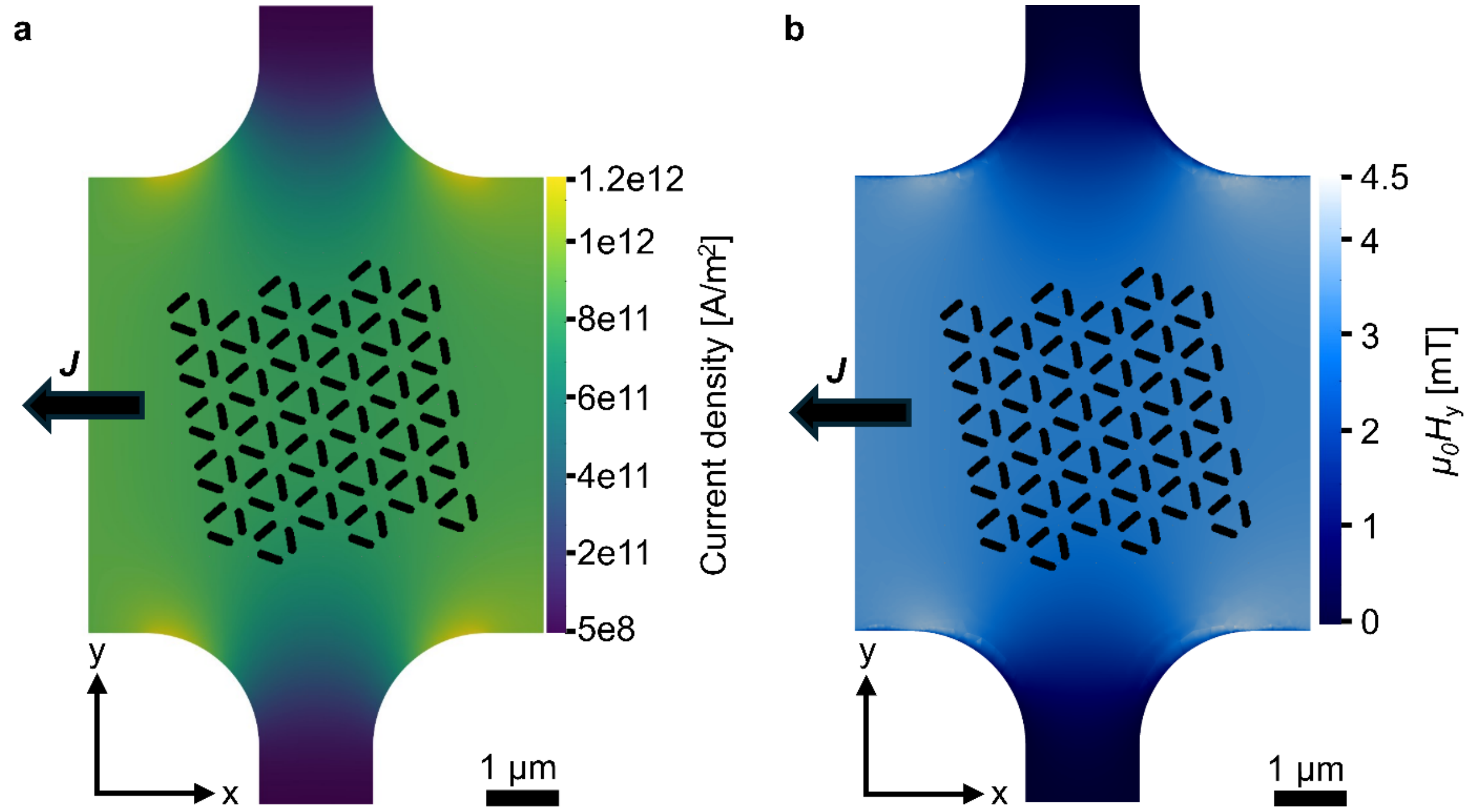

**Supplementary Figure 1: Determination of the current density and magnetic field strength over the Hall bar**. COMSOL finite-element method simulations of the distributions in the Hall bar of (a) the current density $\boldsymbol{J}$ and (b) the Oersted field, computed at a current of 48 mA. The Oersted field is evaluated at a height of 1 nm above the Hall cross surface, corresponding to the mid-plane of the nanomagnets. The position of the triangular ASI is indicated with an overlaid array of black stadium-shaped elements.

## Supplementary Note 2. Micromagnetic simulations

Micromagnetic simulations using MuMax3[1] were carried out to investigate the planar Hall resistance response of the nanomagnets as a function of probing magnetic field and its orientation, presented in lower panels of Supplementary Fig. 2.

In the simulations, the stadium-shaped nanomagnets have dimensions of 350 nm × 100 nm × 2.3 nm, discretized into cells of 4 nm × 4 nm × 2.3 nm. The saturation magnetization and exchange constant were set to $M_s$ = 900 kA/m and $A_{ex}$ = 13 pJ/m, respectively, which are standard simulation parameters for CoFeB[2]. The damping constant was set to $\alpha = 1$ to accelerate convergence to the equilibrium state. The applied magnetic field is swept from −30 mT to +30 mT in steps of 2 mT, with the magnetization allowed to relax at each step. The planar Hall resistance is modeled as proportional to:

$$\begin{aligned} R_H &\propto \sin(2\theta) \\ &\propto \sin\left(2\arctan\left(\frac{M_y}{M_x}\right)\right) \end{aligned}$$

where θ is the polar angle of the nanomagnet magnetization, and $M_x$ and $M_y$ are the $x$ and $y$ components of the magnetization, respectively.

The simulated results agree well with the experimental measurements, as shown by comparing the lower and upper panels of Supplementary Fig. 2.

The simulated magnetization configuration before and after field-induced reversal is shown in Supplementary Fig. 3. In the presence of an applied magnetic field, the magnetization within the nanomagnet adopts a non-uniform configuration, reflecting the competition between the applied field and the shape anisotropy. Reversal occurs abruptly once the applied field exceeds the coercive field, as evidenced by the sharp transition in the simulated $R_H$ versus field curve.

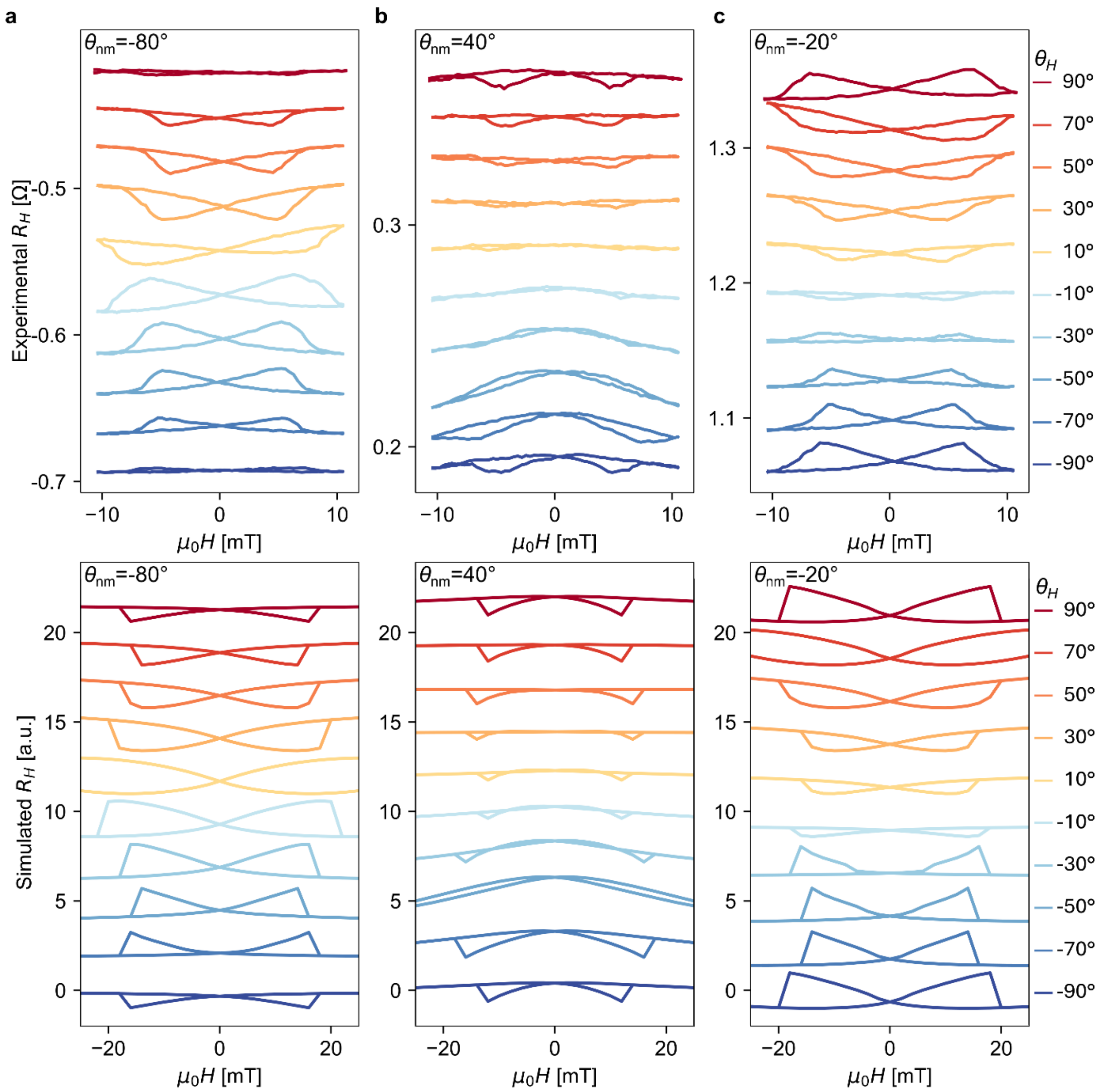

**Supplementary Figure 2: Experimental and simulated planar Hall resistance signal**. Planar Hall resistance $R_H$ as a function of applied magnetic field, measured at various probing field angles $\theta_H$. Each column corresponds to a different nanomagnet orientation: (a) $\theta_{\mathrm{nm}} = -80°$; (b) $\theta_{\mathrm{nm}} = 40°$; (c) $\theta_{\mathrm{nm}} = -20°$. Upper panels are the experimental measurements; lower panels are the corresponding micromagnetic simulation results.

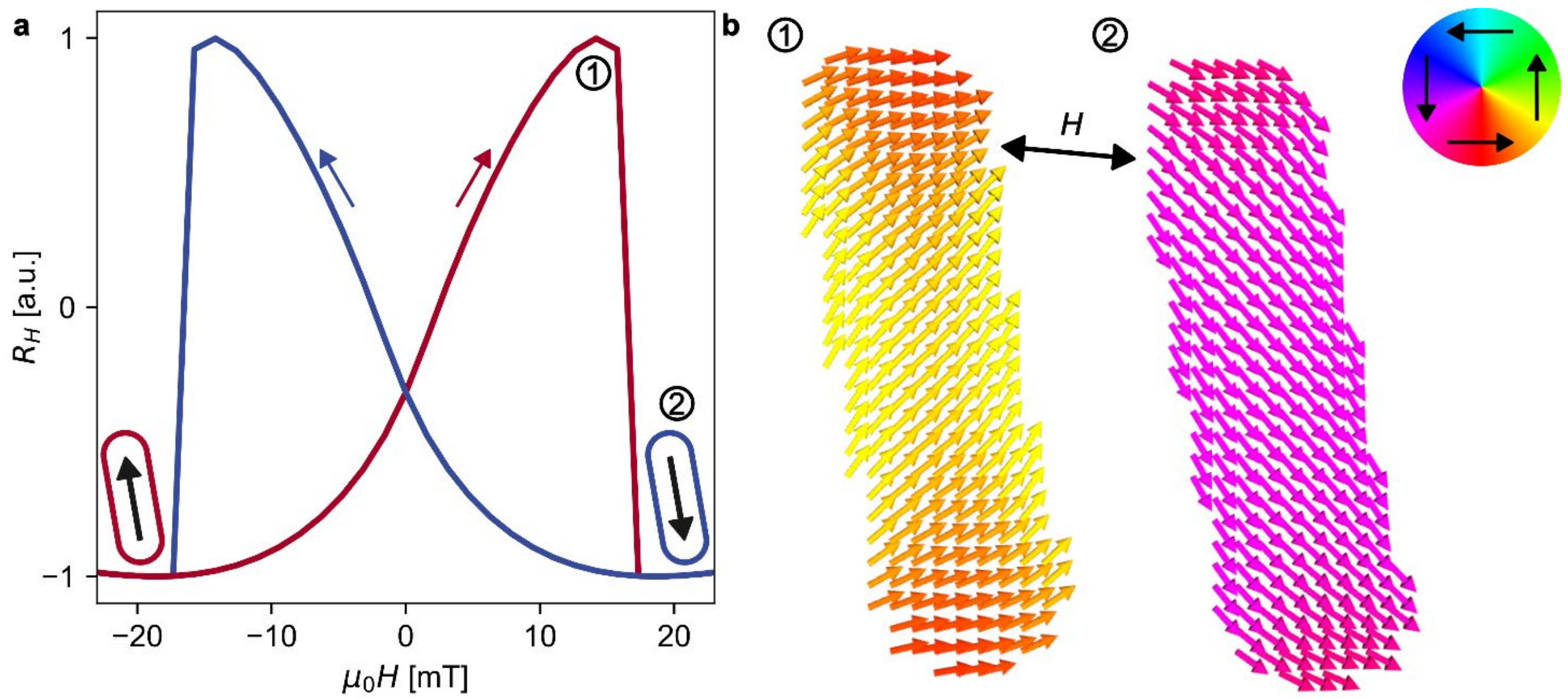

**Supplementary Figure 3: Micromagnetic simulation of field-induced magnetization reversal**. A magnetic field $\mu_0$H is applied at $\theta_H = -10°$ to a nanomagnet oriented at $\theta_{\mathrm{nm}} = -80°$. (a) Simulated planar Hall resistance $R_H$ as a function of applied magnetic field, with the blue and red arrows next to the corresponding curve indicating the field sweep direction. The two branches correspond to the two opposite initial magnetization states of the nanomagnet. (b) Snapshots of the simulated magnetization configuration before ①, and after ②, magnetization reversal, illustrating the field-induced switching of the nanomagnet between its two stable states.

## Supplementary Note 3. Macrospin simulations

Macrospin simulations were performed using the CMTJ package[3] to compute the critical switching current as a function of the ratio of the field-like torque ($\tau_{FL}$) to the damping-like ($\tau_{DL}$) torque, and are presented in Supplementary Fig. 4. The saturation magnetization was set to $M_s$ = 900 kA/m, the nanomagnet thickness to $d$ = 2.3 nm, the spin Hall angle to $\theta_{Hall}$ = 0.25 [4], and the damping constant to $\alpha$ = 0.025, which are standard simulation parameters for CoFeB[2]. The demagnetization matrix $D$ was set to be:

$$D=\begin{bmatrix} 0.0155 & 0 & 0 \\ 0 & 0.0025 & 0 \\ 0 & 0 & 0.9820 \end{bmatrix}$$

with the demagnetization coefficients obtained from Ref. [5]. The damping-like effective field $H_{DL}$ was computed as[1]:

$$\mu_0 H_{\mathrm{DL}} = \frac{\hbar \theta_{\mathrm{Hall}} J}{2 M_s e d}$$

where ħ is the reduced Planck constant, $e$ the electron charge, $M_s$ the saturation magnetization, $d$ the thickness, $\theta_{Hall}$ the spin Hall angle, and $J$ the current density. The simulations were performed by applying a trapezoidal current pulse of 4 ns duration with a rise and fall time of 1 ns and a total simulation time of 20 ns. The current pulse amplitude was incremented in steps of $5 \times 10^9$ A/m$^2$.

With the progressive introduction of a field-like torque component, two distinct switching regimes emerge, characterized by a sudden drop in the critical switching current as the nanomagnet orientation evolves from $\theta_{nm} \rightarrow 0°$ (Type X) to $\theta_{nm} \rightarrow 90°$ (Type Y), as highlighted by the vertical dashed lines in Supplementary Fig. 4. As the field-like torque component increases, the Type X switching regime progressively extends towards 90°, indicating that the field-like torques assist this switching process. As the field-like torque magnitude is increased, the critical switching current for Type X switching (at $\theta_{nm}$ values to the left of the dashed lines) decreases and approaches the critical switching current magnitudes of Type Y switching (at $\theta_{nm}$ values to the right of the dashed lines). Finally, at $\tau_{FL}/\tau_{DL}$ = 0.5, the magnetization switching regime is dominated by the Type X switching and the angular dependence is suppressed. Our experimental data (black points in Supplementary Fig. 4) is therefore best reproduced in the regime where the field-like torque is negligible, that is $\tau_{FL}/\tau_{DL} \rightarrow 0$, consistent with the monotonic increase of the critical switching current as $\theta_{nm} \rightarrow 0°$.

In summary, although field-like torques can reduce the critical switching current for nanomagnets oriented close to $\theta_{nm}$ = 0°, they are detrimental for orientation-selective SOT switching because they introduce a non-monotonic angular dependence of the critical switching current, so that the angular dependence that underpins the selective addressing scheme is lost.

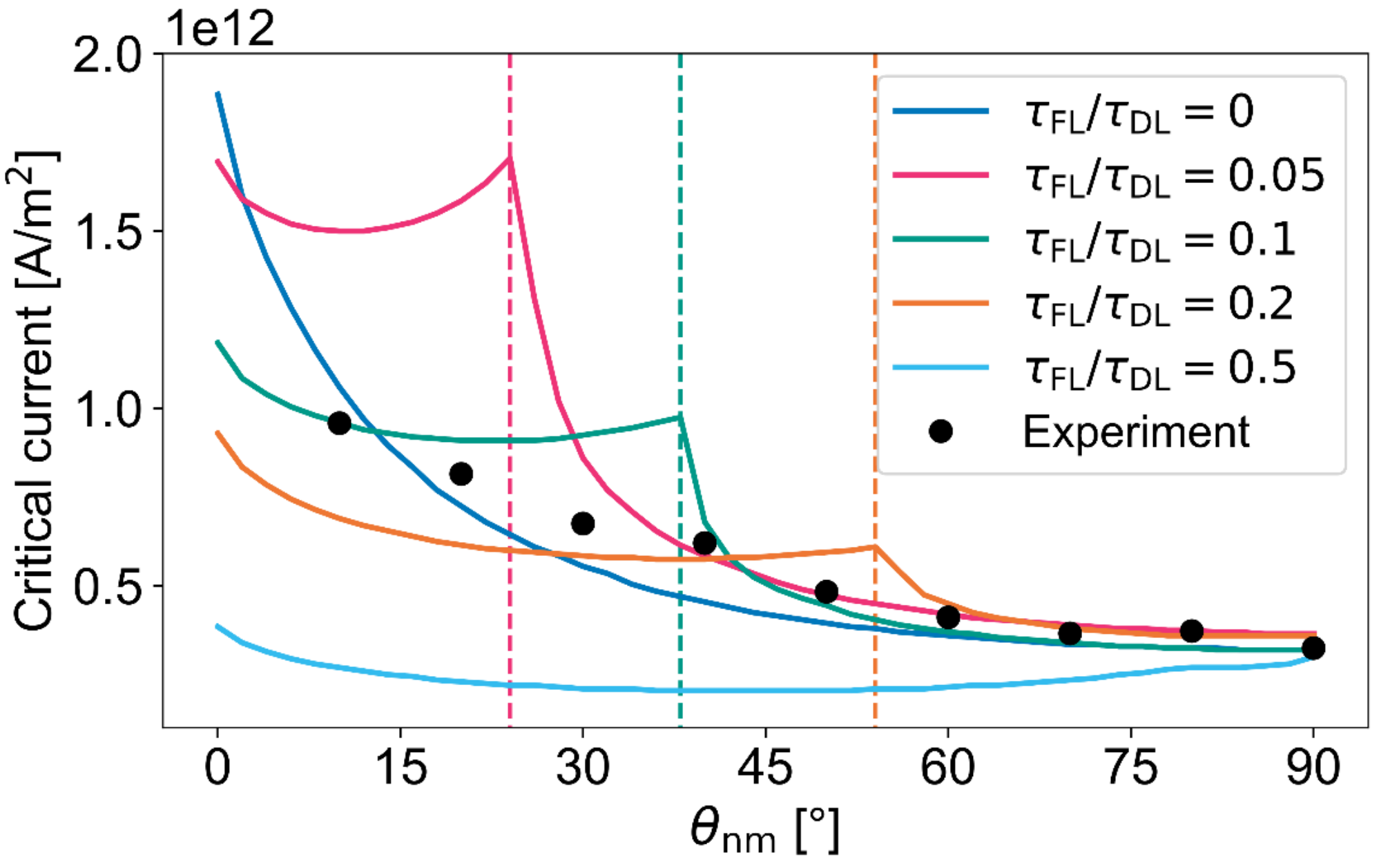

**Supplementary Figure 4: Macrospin simulation of the critical switching current.** Critical switching current as a function of nanomagnet orientation $\theta_{nm}$, computed from macrospin simulations for different ratios of the field-like torque ($\tau_{FL}$) to damping-like torque ($\tau_{DL}$). The vertical dashed lines correspond to the transition from Type X (to the left of the line) to Type Y (to the right of the line) switching for each curve. The experimental data (black dots) is given for comparison.

## Supplementary Note 4. Arrays of nanomagnets with two sublattices

We measured three arrays of nanomagnets with two sublattices, comprising nanomagnets oriented at $\theta_{\mathrm{nm}}$ = -80° and -20°, at -80° and 40°, and at 40° and -20°. Supplementary Fig. 5 shows their SEM images together with the $\Delta R$ versus current loops measured for $\theta_H$ between -90° and 90°, and Supplementary Fig. 6 shows selected loops with the corresponding magnetization states. The nanomagnet array with $\theta_{\mathrm{nm}}$ = -80° and -20° is discussed in the main text (Fig. 3); here we discuss the other two.

For the nanomagnet array comprising the $\theta_{\mathrm{nm}} = -80°$ and $\theta_{\mathrm{nm}} = 40°$ sublattices, presented in Supplementary Fig. 6b, at $\theta_H = 90°$ (top plot), four distinct switching events are observed, confirming that the switching of each sublattice can be unambiguously detected. The angle of $\theta_H = 90°$ was chosen because the $\Delta R$ contribution from each sublattice is equal in amplitude, with $|\Delta R| \approx 5$ mΩ. At $\theta_H = -70°$ (middle plot), the $\Delta R$ contributions from each sublattice are equal in amplitude but opposite in sign when their $y$-magnetization components point towards the same direction. Thus, when the nanomagnets oriented at $\theta_{\mathrm{nm}} = -80°$ and $\theta_{\mathrm{nm}} = 40°$ have both their $y$-magnetization components pointing towards $+y$ (state ①) or $-y$ (state ④), the resulting $\Delta R$ is 0. At $\theta_H = -80°$ (bottom plot), we detect only the magnetization switching of the $\theta_{\mathrm{nm}} = 40°$ sublattice as the $\Delta R$ signal of the $\theta_{\mathrm{nm}}$ = -80° sublattice vanishes.

For the nanomagnet array comprising the $\theta_{\mathrm{nm}} = 40°$ and $\theta_{\mathrm{nm}}$ = -20° sublattices presented in Supplementary Fig. 6c, four distinct switching events are observed at $\theta_H = 90°$ (top plot), highlighted in pink and green for the $\theta_{\mathrm{nm}} = 40°$ and $\theta_{\mathrm{nm}} = -20°$ sublattices, respectively, confirming that the switching of each sublattice can be independently detected. Upon rotating the probing magnetic field by 180° to $\theta_H = -90°$ (middle plot), the $\Delta R$ versus current response is identical in shape but opposite in sign, as expected from the reversal of the probing magnetic field direction, which interchanges the positive and negative $\Delta R$ contributions from each sublattice. At $\theta_H = 40°$ (bottom plot), the $\Delta R$ contribution from the $\theta_{\mathrm{nm}} = 40°$ sublattice vanishes, and only the switching events of the $\theta_{\mathrm{nm}} = -20°$ sublattice are recorded.

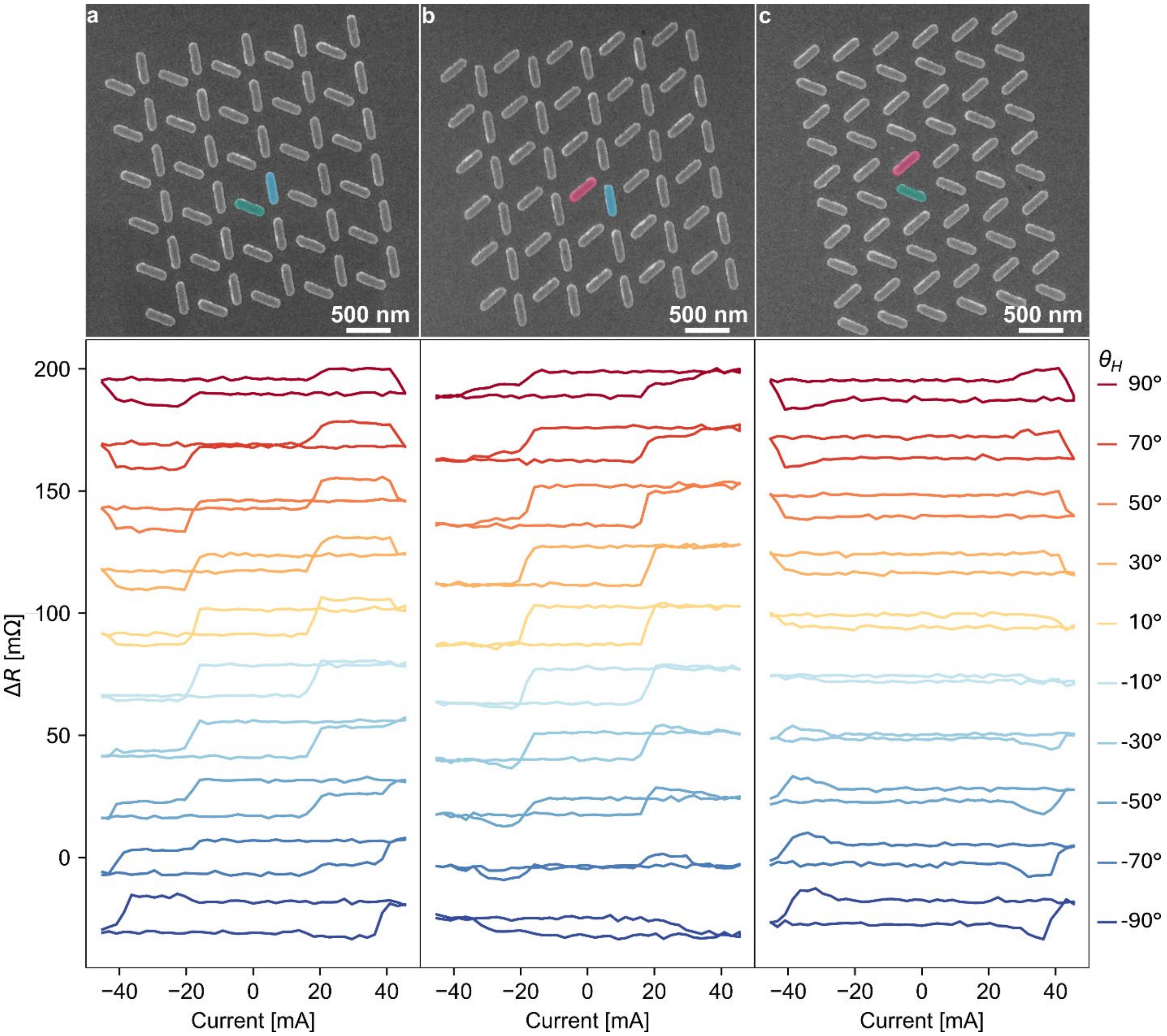

**Supplementary Figure 5: $\Delta R$ of the two-orientation nanomagnet arrays as a function of probing angle**. (a)-(c) top part: SEM images of two-orientation nanomagnet arrays comprising nanomagnets oriented at a) $\theta_{\mathrm{nm}} = -80^\circ$ and $-20^\circ$, (b) $\theta_{\mathrm{nm}} = -80^\circ$ and $40^\circ$, and (c) $\theta_{\mathrm{nm}} = 40^\circ$ and $-20^\circ$. (a)-(c) bottom part: Corresponding $\Delta R$ as a function of applied current pulse amplitude for each configuration, measured at various probing field angles $\theta_H$ ranging from $-90^\circ$ to $90^\circ$.

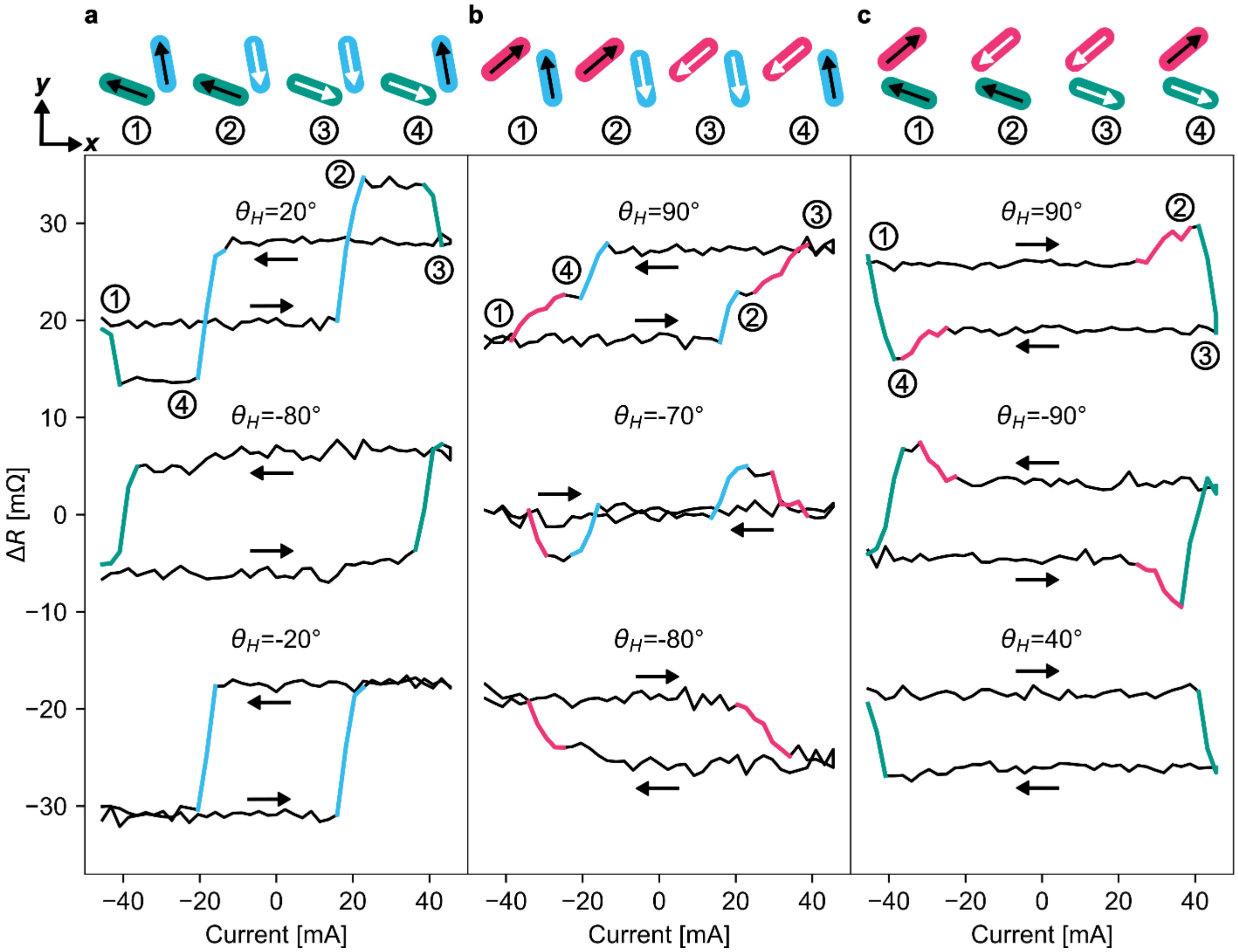

**Supplementary Figure 6: Tunable electrical readout of SOT-induced switching in arrays of nanomagnets with two nanomagnet sublattices**. (a)-(c) $\Delta R$ as a function of applied current pulse amplitude, measured at various $\theta_H$, with the nanomagnets oriented at (a) $\theta_{nm} = -80^{\circ}$ and $-20^{\circ}$, (b) $\theta_{nm} = -80^{\circ}$ and $40^{\circ}$, and (c) $\theta_{nm} = 40^{\circ}$ and $-20^{\circ}$. The switching events of the nanomagnets oriented at $\theta_{nm} = -80^{\circ}$, $40^{\circ}$, and $-20^{\circ}$ are highlighted in blue, pink, and green, respectively. The magnetization state of each two-orientation configuration is schematically illustrated in the initial state and after each switching event. The black and white arrows on the nanomagnets indicate the nanomagnet magnetization pointing towards $+y$ and $-y$, respectively, highlighting the magnetization switching events. The small black arrows next to the curves indicate the current sweep direction.

## Supplementary Note 5. Triangular lattice

In Supplementary Fig. 7, we present the SEM image with the $\Delta R$ versus current loops measured for $\theta_H$ between -90° and 90° for the triangular lattice.

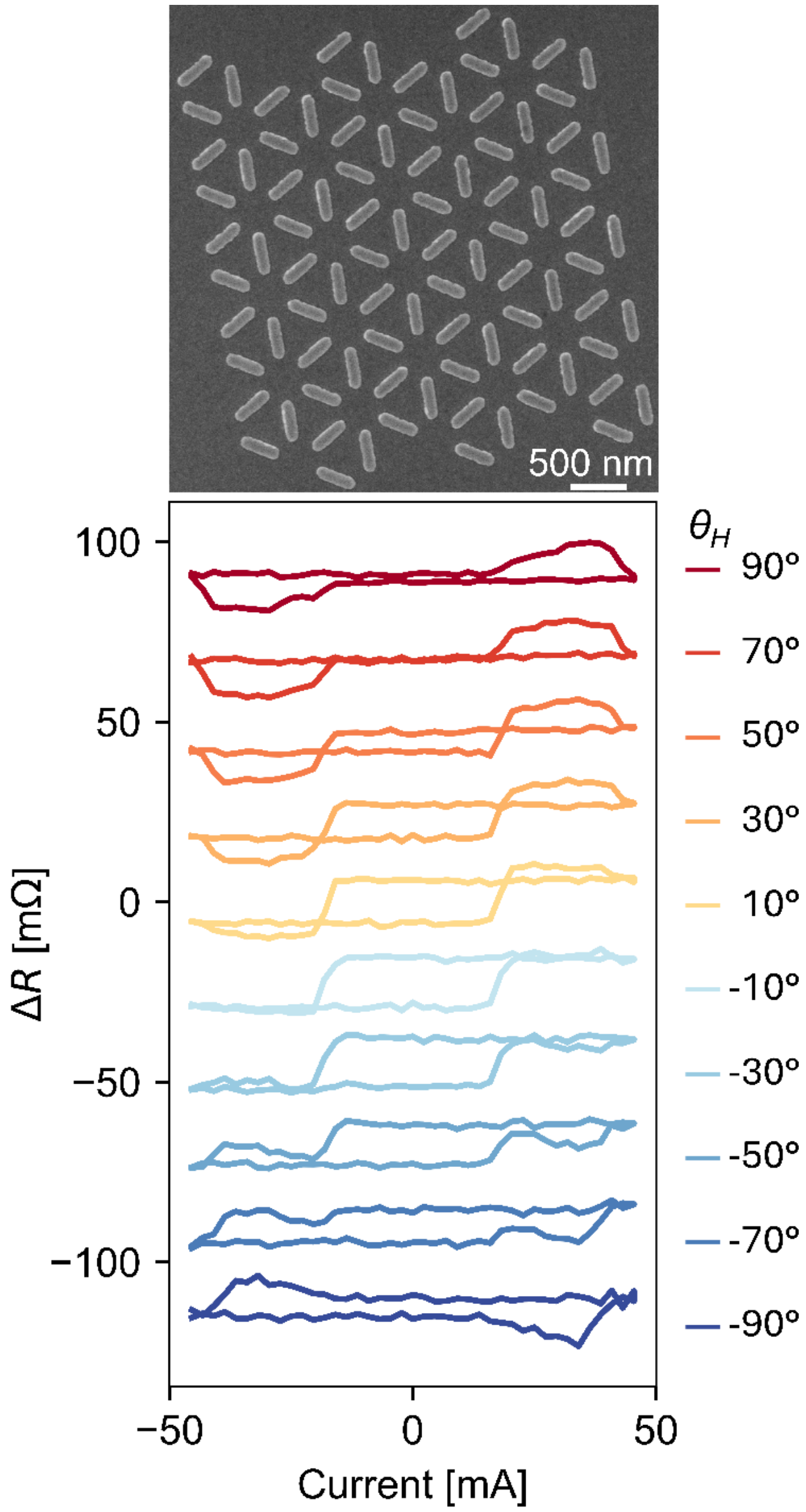

**Supplementary Figure 7: $\Delta R$ of the triangular lattice as a function of probing angle**. Top panel: SEM image of the nanomagnets arranged on a triangular lattice. Bottom panel: $\Delta R$ as a function of applied current pulse amplitude for the triangular lattice, measured at various probing field angles $\theta_H$ ranging from $-90^{\circ}$ to $90^{\circ}$.